\documentclass[12pt]{article}

\advance \textheight by \topskip
\def\R{\mathbb R}

\def\N{\mathbb N}
\def\Z{\mathbb Z}

\def\asl{\mathfrak{sl}}

\def\cF{\mathcal F}
\def\cH{\mathcal H}

\def\I{\hbox{{1\hskip -5.8pt 1}\hskip -3.35pt I}}

\def\pd{\partial}

\def\mop #1{\mathop{\sf #1}\nolimits}

\usepackage{amsmath,amssymb,euscript,graphicx}

\begin{document}
\title{{\bf
Quasi-exact solvability and intertwining relations}}
\author{{\sf Sergey Klishevich}\thanks{E-mail:
klishevich@ihep.ru}
\\{\small {\it Institute for High Energy Physics, Protvino,
Russia}}}
\date{}
\maketitle

\vskip-1.0cm

\begin{abstract}
We emphasize intertwining relations as a universal tool in
constructing one-dimensional quasi-exactly solvable
operators and offer their possible generalization to the
multidimensional case. Considered examples include all
quasi-exactly solvable operators with invariant subspaces of
monomials. We show that the simplest case of generalized
intertwining relations allows to naturally relate
quasi-exactly solvable operators with two invariant monomial
subspaces to a nonlinear parasupersymmetry of second order.
Quantum-mechanical systems with linear and nonlinear
supersymmetry are discussed from the viewpoint of
quasi-exact solvability. We construct such a general system
with a cubic supersymmetry and argue that quantum-mechanical
systems with nonlinear supersymmetry of fourth order and
higher are generally not quasi-exactly solvable. Besides, we
construct two examples of quasi-exactly solvable operators
with invariant subspaces which cannot be reduced to monomial
spaces.
\end{abstract}
\newpage

\section{Introduction}

In the eighties a new class of quantum-mechanical problems
was discovered. Spectra of such systems can be partially
found by an algebraic procedure, therefore they were called
\textit{quasi-exactly solvable} systems. In this sense these
problems occupies an intermediate position between exactly
solvable and unsolvable systems. There were elaborated
several approaches to constructing quasi-exactly solvable
quantal systems. Among them the Lie-algebraic construction
\cite{qes} and analytic approach \cite{ushv} are most
common. In the Lie-algebraic approach a quasi-exactly
solvable Hamiltonian is a polynomial of differential
operators which form a finite-dimensional representation of
a Lie algebra. This automatically guaranties the quasi-exact
solvability of the Hamiltonian. In one-dimensional quantal
problems the corresponding algebra is $\asl(2,\R)$. Later on
such 1D systems were generalized to quasi-exactly solvable
operators with invariant monomial subspaces \cite{turb&post}
(for recent discussions see Refs. \cite{kamran03,
kamran04}).

Recently it was revealed nontrivial relations between
quasi-exactly solvable systems and nonlinear supersymmetry
\cite{nsusy0, nsusy1, nsusy2, nsusy3} which is a
generalization of the linear supersymmetric quantum
mechanics \cite{nicolai, witten, cooper} to the case of
nonlinear (polynomial) superalgebras \cite{andrian, plyush}.
The correspondence between the Lie-algebraic scheme based on
the algebra $\asl(2,\R)$ and supersymmetric systems with
nonlinear superalgebras was shown in Ref. \cite{nsusy2}
while supersymmetric systems corresponding to the
quasi-exactly solvable operators with invariant monomial
subspaces were constructed in Ref. \cite{gonzalez&tanaka}.

Intertwining relations underline the standard realization of
the quantum-mechanical supersymmetry both linear and
nonlinear. The only difference of the nonlinear case is that
the corresponding intertwining differential operators are of
higher orders. Intertwining relations are corner stone of
various approaches such as shape invariance to constructing
exactly solvable systems \cite{cooper, A-related}. In this
paper we emphasize the role of intertwining relations in
constructing quasi-exactly solvable operators and offer a
generalization of the construction to the multidimensional
case.

The paper is organized as follows. In the next section we
discuss the relation of intertwining relations to
quasi-exact solvability of one-dimensional operators.
Besides we offer a natural algebraic generalization of
intertwining relations to the multidimensional case. In the
section \ref{1D} we consider definition and some aspects of
the one-dimensional quasi-exactly solvable operators. In the
section \ref{sl2sec} the results of the Lie-algebraic scheme
based on the algebra $\asl(2,\R)$ are reproduced in terms of
the method based on the intertwining relations. In the
section \ref{monomials} we discuss application of the method
to quasi-exactly solvable operators with invariant monomial
spaces. An intimate relation of quasi-exactly solvable
operators with two invariant monomial subspaces to a
nonlinear parasupersymmetry of second order is outlined in
the section \ref{2spaces}. The section \ref{sqm} is devoted
to investigation of one-dimensional quantum-mechanical
systems with linear and nonlinear supersymmetry from the
viewpoint of quasi-exact solvability. In this section we
construct a general quasi-exactly solvable operator with
three-dimensional invariant subspace, which is related to a
cubic supersymmetry. Besides, using the method based on
intertwining relations two new examples of quasi-exactly
solvable operators are constructed. The corresponding
invariant subspaces do not belong to spaces with a monomial
basis but can be considered as their one-parametric
deformation. The conclusion discusses and summarizes the
obtained results.

\section{Intertwining relations and quasi-exact solvability}
\label{ir&qes}

In a general sense a differential operator is called
quasi-exactly solvable if it has a finite-dimensional
invariant subspace. If this subspace is known explicitly the
corresponding part of spectral problem can be solved
algebraically.

Let us consider the following intertwining relations
\begin{align}\label{Dt}
  AH_0&=H_1A, & A^\dag H_1&=H_0A^\dag,
\end{align}
where $H_0$, $H_1$ and $A$ are some operators on a Hilbert
space $\cH$ and $^\dag$ denotes an involution, e.g. a
hermitian conjugation. As is well known the intertwining
operator $A$ relates the spectra of the operators $H_1$ and
$H_0$:
\begin{align*}
  \sigma(H_0)&\stackrel{A}{\longrightarrow}\sigma(H_1),&
  \text{i.e. }&&\tilde\psi_E &= A\psi_E
\end{align*}
for $\psi_E\notin \mop{Ker}A=\{\psi\in\cH\ |\ A\psi=0\}$,
where $E$ stands for an eigenvalue. Besides, from the
intertwining relations \eqref{Dt} it follows that if
$\mop{Ker}A$ is not empty then it is a nontrivial invariant
subspace of the operator $H_0$:
$$
 \mop{Ker}A\stackrel{H_0}\longrightarrow\mop{Ker}A.
$$
This property is very interesting from the viewpoint of
quasi-exact solvability. Indeed, if $\mop{Ker}A$ is a
finite-dimensional space then the operator $H_0$ is
quasi-exactly solvable. Hence, for a given operator $A$ with
a finite number of zero-modes the intertwining relations
\eqref{Dt} provide a constructive way to build corresponding
quasi-exactly solvable operator $H_0$.

When the operators $H_0$ and $H_1$ are hermitian a
superalgebra\footnote{The operators also can be
pseudo-hermitian and in this case one can associate a
pseudo-supersymmetry \cite{pseudo-susy, nsusy3} with them.}
can be associated with the intertwining relations
\eqref{Dt}. Indeed, introducing the pair of odd variables
$\theta$, $\theta^\dag$, $\{\theta,\,\theta^\dag\}=1$, one
can construct the supercharges and super-Hamiltonian:
\begin{align}\label{susy}
  Q&=A\theta,& Q^\dag&=A^\dag\theta^\dag,&
 \hat H&=H_0\theta\theta^\dag + H_1\theta^\dag\theta.
\end{align}
The supercharges commute with the super-Hamiltonian due to
the intertwining relations \eqref{Dt}. Operators $\hat H$,
$Q$, $Q^\dag$ generate a linear superalgebra \cite{nicolai,
witten} when $A$ is a linear differential operator. If $A$
is a higher order differential operator the corresponding
superalgebra is of nonlinear (polynomial) form \cite{nsusy0,
nsusy1, nsusy2, nsusy3, nsusy4}.

The intertwining relations \eqref{Dt} are very useful for
constructing one-dimensional quasi-exactly solvable
operators since $A$ being a one-dimensional differential
operator always has a finite number of zero-modes. In
multidimensional problems it is not the case and the scheme
based on the intertwining relations should be modified in
appropriate way. To this end we offer the following
generalization of the intertwining relations (\ref{Dt}):
\begin{align}\label{gDt}
 A_iH_0&=\sum_{j=1}^MH_{ij}A_j, &
   \text{for \hskip 8mm} i=1,\,\ldots,M.
\end{align}
In this case if the operator-valued matrix $H_{ij}$ is
``irreducible'', i.e. it is not of a block-diagonal form,
and $H_{ii}\neq 0$ for all $i$,
the space
\begin{equation}\label{cF}
 \cF(H_0)=\bigcap_{i=1}^M\mop{Ker}A_i
\end{equation}
is invariant subspace of the operator $H_0$. In general,
depending on the particular form of the operator-valued
matrix $H_{ij}$ there can be some other invariant subspaces
of $H_0$. For example, consider following two cases:
\begin{align}\label{msp}
 A_1H_0&=H_1A_1,&  A_2H_0&=H_2A_2,\\
\intertext{and}
 A_1H_0&=H_{11}A_1,&  A_2H_0&=H_{21}A_1+H_{22}A_2.
\end{align}
In the former case if the intersection of $\mop{Ker}A_i$ is
not empty there are three invariant subspaces of the
operator $H_0$:
$\mop{Ker}A_i\setminus(\mop{Ker}A_1\cap\mop{Ker}A_2)$,
$i=1$, $2$ and $\mop{Ker}A_1\cap\mop{Ker}A_2$. In the latter
there are two invariant subspaces:
$\mop{Ker}A_1\setminus(\mop{Ker}A_1\cap\mop{Ker}A_2)$ and
$\mop{Ker}A_1\cap\mop{Ker}A_2$. More detailed analysis of
the generalized intertwining relations \eqref{gDt} in the
context of invariant subspaces of the operator $H_0$ is out
of the scope of the present paper and will be given
elsewhere. We just would like to add that although in the
multidimensional case any space of zero-modes $\mop{Ker}A_i$
generally has infinite dimension the invariant subspace
\eqref{cF} can be finite-dimensional. Therefore the
relations \eqref{gDt} can be used for constructing
multidimensional quasi-exactly solvable systems.

\section{Quasi-exactly solvable operators in one dimension}
\label{1D}

Let us consider the most general differential operator of
second order:
\begin{equation}\label{H}
 H={}-P(z)\pd_z^2+Q(z)\pd_z+R(z),
\end{equation}
where we assume that $z\in I=\{z\in\R\ |\ P(z)>0\}$,
$\pd_z=d/dz$, $P(z)$, $Q(z)$ and $R(z)$ are some real
functions. Formally, the operator $H$ is hermitian with
respect to the following inner product:
\begin{align}\label{sp}
 \langle\phi,\,\psi\rangle_\rho&=
   \int_I\phi(z)\psi(z)\rho(z)dz, &
   \rho(z)&=P(z)^{-1}\exp
      \left(-\int\limits_c^z\frac{Q(y)}{P(y)}\,dy\right),
\end{align}
where $\phi(z)$ and $\psi(z)$ are real functions. The
generalization to the supersymmetric case is
straightforward. The wave functions become two component
functions and the weight function $\rho(z)$ should be
changed for a diagonal matrix defined from hermicity of the
Hamiltonians $H_0$, $H_1$.

We require for the Hamiltonian \eqref{H} to be a covariant
operator with respect to coordinate transformations.
Therefore, under the change of variable $z\to x(z)$ the
functions $P(z)$, $Q(z)$ and $R(z)$ transform as
\begin{align}\label{dm}
 \tilde P(x(z))&=(\partial_zx(z))^2P(z),\notag\\
 \tilde Q(x(z))&=\partial_zx(z)Q(z)-\partial_z^2x(z)P(z),
 \\\notag
 \tilde R(x(z))&=R(z).
\end{align}
When matrix elements of the Hamiltonian \eqref{H} are
invariant under the coordinate transformations:
$$
 \langle\tilde\psi,\tilde H\tilde\psi\rangle_{\tilde\rho}=
 \langle\psi,H\psi\rangle_\rho,
$$
where we assume that the wave functions are normalized and
transform as scalar fields, $\tilde\psi(x(z))=\psi(z)$,
while the weight function $\rho$ transforms according to its
definition \eqref{sp}.

Besides, the system \eqref{H}, \eqref{sp} has another local
symmetry. The matrix elements are invariant under the
so-called ``gauge'' transformations:
\begin{align}\label{gt0}
  \tilde\psi(z)&=e^{\alpha(z)}\psi(z),&
  \tilde H&=e^{\alpha(z)}He^{-\alpha(z)}.
\end{align}
In terms of the coefficient functions of the Hamiltonian it
reads as
\begin{align}\label{gt}
 \tilde P(z)&=P(z),\notag\\
 \tilde Q(z)&=Q(z)+2\pd_z\alpha(z)P(z),\\\notag
 \tilde R(z)&=R(z)-\pd_z\alpha(z)Q(z)
   + \left(\pd_z^2\alpha(z)-(\pd_z\alpha(z))^2\right)P(z).
\end{align}
Form the definition \eqref{sp} one can deduce the
transformation for $\rho(z)$:
$$
  \tilde\rho(z)=\lambda e^{-2\alpha(z)}\rho(z).
$$
The constant factor $\lambda$ depends on the integration
constant $c$ in the definition of $\rho(z)$ \eqref{sp} but
it is irrelevant until we use normalized wave functions.

The Hamiltonian \eqref{H} and the inner product \eqref{sp}
can be represented in an explicitly covariant form with
respect to the transformations \eqref{dm} and \eqref{gt}. To
this end, consider the following operator:
\begin{align}\label{cH}
 H&={}-\frac 1{\sqrt{g}}\nabla_z(\sqrt{g}g^{zz}\nabla_z)
      + V(z),
\end{align}
where the metric is given by the quadratic form
$ds^2=g_{zz}(z)dz\otimes dz$,
$g=\mop{det}\|g_{zz}\|=g_{zz}$, $g^{zz}g_{zz}=1$ and $V(z)$
is a scalar function. $\nabla_z=\pd_z-A_z$, where $A_z$ is
component of the one-form ${\cal A}=A_z(z)dz$. The
definitions of the metric and the one-form fix how their
components change under the coordinate transformations while
we assume that under the "gauge" transformations \eqref{gt0}
the 1D vector potential transforms as an Abelian connection
\begin{equation}\label{gtA}
 \tilde A_z=A_z + \pd_z\alpha.
\end{equation}
Therefore, the transformations \eqref{gt} are true gauge
transformations and $\nabla_z$ is the corresponding
covariant derivative. Hence, the Hamiltonian \eqref{cH} has
an explicitly covariant form.

The appropriate invariant inner product has the form
\begin{equation}\label{csp}
  \langle\phi,\,\psi\rangle=
   \int\phi(z)\psi(z)e^{-2\int^z\!\!\mathcal A}\sqrt{g}dz,
\end{equation}
where the exponential factor provides the invariance of the
inner product with respect to the gauge transformations
\eqref{gt0}, \eqref{gtA}.

The link between the covariant formulation \eqref{cH},
\eqref{csp} is given by the relations
\begin{align*}
 P(z)&=g^{zz},&
 Q(z)&=2g^{zz}\left(A_z+\Gamma^z_{zz}\right),&
 R(z)&=V(z) - A^zA_z + g^{zz}\nabla_zA_z,
\end{align*}
where $\Gamma^z_{zz}=\pd_z\log\sqrt g$. So, there is a
one-to-one correspondence between the coefficient functions
of the Hamiltonian \eqref{H} and $g_{zz}$, $A_z$, $V(z)$.

Now let us discuss one-dimensional quasi-exactly solvable
operators. If the operator \eqref{H} is quasi-exactly
solvable it has a finite-dimensional invariant subspace,
say,
\begin{align}\label{space0}
 \cF(H)&=\mop{span}\{f_1(z),\,f_2(z),\,\ldots,\,f_n(z)\},
\end{align}
where linearly independent real functions $f_i(z)$ form a
basis of the subspace. One can think that the functions are
ordered by number of their nodes. Hereinafter we will denote
a finite-dimensional invariant subspace of an operator
$\mathcal O$ as $\cF(\mathcal O)$.

In principle one can extend this definition and treat an
operator as quasi-exactly solvable one if a part of its
spectrum (such part can be infinite-dimensional) can be
found algebraically. Here we discuss only quasi-exactly
solvable operators with finite-dimensional invariant
subspace.

Generally, by a change of the coordinate and the gauge
transformation (\ref{gt0}) one can reduce the invariant
subspace (\ref{space0}) to the following form
\cite{turb&post}:
\begin{align}\label{space1}
 \cF(H)&=\mop{span}\{1,z,f_3(z),\,f_4(z),\,\ldots,\,
  f_n(z)\},
\end{align}
From this form it is obvious that the invariant subspaces of
dimensionality 1 and 2 are special cases and require a
special treatment.

At the end of this section we would like to briefly discuss
the general structure of quasi-exactly solvable operator. As
was proved in Ref. \cite{kamran99} every linear or nonlinear
quasi-exactly solvable operator with a quite general
finite-dimensional invariant subspace \eqref{space0} can be
represented in the following form:
\begin{equation}\label{gen_qes}
 T_{qes}[\psi]=
    \sum_{i=1}^{n}f_i(z)\,F_i(L_1[\psi],\,\ldots,L_n[\psi])
   +\sum_{\alpha=1}^{l}A_\alpha[\psi]\,\mathcal
   O_\alpha[\psi],
\end{equation}
where the $F_i\in\mathcal C^\omega(\mathbb R^n)$,
$\mathcal O_\alpha$ are arbitrary operators, $A_\alpha$ are
annihilating operators (annihilators) of the space
\eqref{space0}, i.e. $A_\alpha[\psi]=0$ for all
$\psi\in\cF(T_{qes})$, and $L_i$ form a basis of the space
of affine annihilators, i.e. $L_i[\psi]=const\in\mathbb R$
for all $\psi\in\cF(T_{qes})$.

\section{Intertwining relations and the $\asl(2,\R)$-based
scheme}
\label{sl2sec}

In this section we consider application of the intertwining
relations \eqref{Dt} to the well-known case of quasi-exactly
solvable operators corresponding to the Lie-algebraic scheme
based on the algebra $\asl(2,\R)$.

The Hamiltonians $H_0$ and $H_1$ \eqref{Dt} are taken in the
most general form
\begin{equation}\label{H_i}
 H_i={}-P_i(z)\pd_z^2+Q_i(z)\pd_z+R_i(z),
\end{equation}
where $P_i(z)$, $Q_i(z)$ and $R_i(z)$ are arbitrary real
functions. Here we treat the intertwining operator $A$ as an
annihilation operator of the corresponding invariant
monomial subspace. Hence, the ``bosonic'' Hamiltonian $H_0$
is a candidate for a quasi-exactly solvable operator while
the operator $H_1$ is its supersymmetric partner in the
sense of the supersymmetric construction \eqref{Dt},
\eqref{susy}.

In the case of the $\asl(2,\R)$-based scheme the invariant
subspace can always be reduced to the form
\begin{equation}\label{Fsl2}
  \cF_n=\mop{span}
     \bigl\{1,z,z^2,\ldots, z^{n-1}\bigr\},
\end{equation}
where $n\in\N$.
It is the representation space of the algebra $\asl(2,\R)$
with generators given by the differential operators
\begin{align}\label{sl2}
 J^+_n&=z^2\pd_z - \left(n-1\right)z, &
 J^0_n&=z\pd_z - \frac{n-1}2, &
 J^-_n&=\pd_z.
\end{align}
Here we assumed that $n>2$ since the cases $n=1,\, 2$ should
be analyzed separately.

In the standard approach any quasi-exactly solvable operator
is taken as a polynomial of the generators \eqref{sl2}.
However we will obtain the general form of such an operator
using the strategy based on the intertwining relations
\eqref{Dt}. This method is more general then the
Lie-algebraic approach since, as we will see later, it is
equally good for constructing any one-dimensional
quasi-exactly solvable operators.

It is very easy to see that an annihilator of the monomial
subspace \eqref{Fsl2} can be represented in the following
simple form:
\begin{equation}\label{A_n}
 A_n=\pd_z^n.
\end{equation}
If we find the operators $H_0$, $H_1$ and $A_n$ which obey
the intertwining relations \eqref{Dt} then $H_0$ is a
quasi-exactly solvable operator with the finite-dimensional
invariant subspace $\mop{Ker}A_n=\cF_n$. The question on the
normalizability of such states we leave aside in this paper.

Substituting the operators \eqref{H_i} and \eqref{A_n} into
the intertwining relations (\ref{Dt}) one gets a system of
differential equations for the functions $P_i(z)$, $Q_i(z)$,
and $R_i(z)$, $i=1,\,2$. This system can be divided into two
parts. The fist one can be solved algebraically and allows
to express the coefficient functions of $H_1$ in terms of
those of $H_0$:
\begin{eqnarray*}
 &&P_1(z)=P_0(z), \label{eq1}\\[2mm]
 &&Q_1(z)=Q_0(z)-nP_0'(z), \label{eq2}\\\label{eq3}
 &&R_1(z)=R_0(z)+nQ_0'(z)-\frac{n(n-1)}2P_0''(z).
\end{eqnarray*}
The second part is an overdetermined system of differential
equations for the functions $P_0(z)$, $Q_0(z)$ and $R_0(z)$:
$$
 m(m-1)P_0^{(n-m+2)}\!(z)-m(n-m+2)Q_0^{(n-m+1)}\!(z)
   - (n-m+1)(n-m+2)R_0^{(n-m)}\!(z)=0,
$$
where $m=0,\,\ldots, n-1$. The equations with
$m\in\{n-3,\,n-2,\,n-1\}$ are independent and give the
following 9-parametric solution:
\begin{eqnarray}
 P_0(z)&=&\sum_{m=0}^4p_mz^m,\notag\\\label{sl2}
 Q_0(z)&=&\sum_{m=0}^2q_mz^m + \frac{n-2}2P_0'(z),\\\notag
 R_0(z)&=&\frac{(n - 1)(n - 2)}6P_0''(z)
 -\frac{n-1}2Q_0'(z)+v,
\end{eqnarray}
where $p_m$, $q_m$, and $v\in\R$.
This solution satisfies the complete overdetermined system
of differential equations. The second order operator $H_0$
with the coefficient functions \eqref{sl2} has the same form
as in the Lie-algebraic scheme based on the algebra
$\asl(2,\R)$.

It is worth noting that the procedure of solving such a
system of differential equations is so simple only in the
case \eqref{Fsl2}.

\section{Generalized quasi-exactly solvable systems with
invariant monomial subspaces}
\label{monomials}

A generalization of one-dimensional quasi-exactly solvable
systems corresponding to the $\asl(2,\R)$-based approach was
investigated for the cases of invariant subspaces with a
monomial basis \cite{turb&post} (for more recent discussion,
see Refs. \cite{kamran03, kamran04}). In this section we
demonstrate that the approach to quasi-exact solvability
based on the intertwining relations \eqref{Dt} can be
applied to such quasi-exactly solvable systems as well.

Now let us discuss how one can deal with a general monomial
invariant subspace using the intertwining relations
\eqref{Dt}. An $n$-dimensional space with monomial basis
functions can be represented as
\begin{equation}\label{mon_sp}
 \cF=\mop{span}
  \bigl\{z^{m_1},\,z^{m_2},\ldots,\, z^{m_n}\bigr\}.
\end{equation}
One can consider the parameters $m_i$ as natural numbers but
for the method based on the intertwining relations it does
not matter and $m_i$ can be treated as real numbers.

The annihilator of the space \eqref{mon_sp} can be taken in
the form
\begin{align}\label{A_mon}
 A_n&=\prod_{i=1}^n(z\pd_z-m_i).
\end{align}
The intertwining relations \eqref{Dt} with the intertwining
operator $A_n$ lead to the overdetermined system of
differential equations on coefficient functions of the
operators $H_0$ and $H_1$. The particular solution of the
system depends on the parameters $m_i$ and order of the
operators $H_0$ and $H_1$. A solution always exists since
$H_0$ taken as a polynomial of the operator $z\pd_z$ leaves
the subspace \eqref{mon_sp} invariant and the form of
coefficient functions of the operator $H_1$ can be found
from the intertwining relations algebraically. We are
interested in nontrivial solutions for the differential
operators $H_0$ and $H_1$ of second order.

In terms of the new variable $\xi=\log z$ one has
$z\pd_z=\pd_\xi$ and the annihilator \eqref{A_mon} becomes a
differential operator with constant coefficients. As a
consequence, the intertwining relations \eqref{Dt} lead to a
system of differential equations with constant coefficients
for the coefficient functions of the Hamiltonians $H_0$ and
$H_1$. A system of this kind can be solved by a purely
algebraic procedure. Therefore the resulting Hamiltonian
$H_0$ is a quasi-exactly solvable operator with the
invariant subspace \eqref{mon_sp}.

Here we will not elaborate this procedure in any detail.
Linear differential operators with finite-dimensional
invariant subspaces of monomials were investigated in Ref.
\cite{turb&post}\footnote{More recent discussion on the
subject can be found in Refs. \cite{kamran03, kamran04}.},
where a classification of second order operators with
invariant subspaces of the form \eqref{mon_sp} was given.

We give here the classification of such finite-dimensional
invariant subspaces with basis of monomial functions without
any explicit form of the corresponding operators. The
form of the operators can be found in Refs. \cite{turb&post,
kamran03, kamran04}. Invariant subspaces of monomials for
nontrivial second order differential operators\footnote{Here
the trivial operator is a polynomial in $z\pd_z$.} can be
divided into two classes. The fist one comprises the
following spaces with a monomial basis of dimension 3 and 4:
\begin{align*}
 \cF_3&=\mop{span}\bigl\{1,\,z,\,z^m\bigr\},&
 \cF_4&=\mop{span}\bigl\{1,\,z,\,z^m,\,z^{2m}\bigr\}.
\end{align*}
The parameter $m$ can be treated as real number. The second
class is represented by series of spaces of monomials. The
first series is that corresponding to the Lie-algebraic
approach \eqref{Fsl2}. The next series is
\begin{align}\label{Fgap}
  \cF&=\mop{span}
 \bigl\{1,\,z,\,z^2,\,\ldots,\,z^{n-2},\,z^n\bigr\},
\end{align}
where $n\in\Z$. The last series is a direct sum of two
subspaces:
\begin{align}\label{2sectors}
 \cF&=\mop{span}\bigl\{
       1,\,z,\,z^2,\,\ldots,\,z^{n-1}
                \bigr\}\oplus
      \mop{span}\bigl\{
       z^m,\,z^{m+1},\,z^{m+2},\,\ldots,\,z^{m+k-1}
                \bigr\}=\cF_n\oplus z^m\cF_k,
\end{align}
where $n,\,k\in\N$ and $m\in\R$.

Up to a change of variable and gauge transformations these
spaces exhaust the list of invariant monomial subspaces of
quasi-exactly solvable differential operators of second
order.

All these cases can be reproduced in terms of the usual
intertwining relations \eqref{Dt}. This method allows to
automatically construct the associated supersymmetric
systems with nonlinear superalgebras (details on the
construction can be found in Ref. \cite{gonzalez&tanaka}).
In the next section we will show that using the simplest
generalized intertwining relations \eqref{msp} a nonlinear
parasupersymmetry can be naturally associated with
quasi-exactly solvable operators invariant on the monomial
spaces \eqref{2sectors}.

\section{Systems with two invariant subspaces and nonlinear
parasupersymmetry}
\label{2spaces}

In quasi-exactly solvable systems with two invariant
subspaces \eqref{2sectors} one can associate an annihilating
operator with each subspace. Consider the following two
operators:
\begin{align}\label{}
 A_1&=\pd_z^n, & A_2&=\pd_z^kz^{-m}.
\end{align}
Their spaces of zero-modes are
\begin{align}
 \mop{Ker}A_1&=\cF_n,&\mop{Ker}A_2&=z^m\cF_k.
\end{align}
If the Hamiltonian $H_0$ is a second order differential
operator and obeys the generalized intertwining relations
\eqref{msp} for some operators $H_1$ and $H_2$ then it is a
quasi-exactly solvable operator with the invariant subspace
\eqref{2sectors}
$$
 \cF(H)=\mop{Ker}A_1\oplus\mop{Ker}A_2=\cF_n\oplus z^m\cF_k.
$$

From the first relation in \eqref{msp} it follows that the
Hamiltonian $H_0$ is a quasi-exactly solvable operator
corresponding to the $\asl(2,\R)$-based scheme. The second
relation from \eqref{msp} can be considered as a constraint
on this quasi-exactly solvable operator. Thus, without
explicit calculations we came to the same conclusion as made
in Ref. \cite{kamran03}. We can even say something more
about this system. It is known that the generalized
intertwining relations \eqref{msp} underlay a
parasupersymmetric system \cite{rubakov}. Indeed, from the
relations \eqref{msp} one concludes
\begin{align*}
 A_i^\dag A_i&=P_i(H_0),& A_iA_i^\dag&=P_i(H_{i}),
\end{align*}
where $i=1,2$, $P_1(.)$ is a polynomial of order $n$ and
$P_2(.)$ is a polynomial of order $k$. Here the hermitian
conjugation is understood with respect to the quite
straightforward generalization of the inner product
\eqref{sp} to the case of 3-component wave functions and
$\rho$ changed for a $3\times 3$ diagonal matrix defined
from hermicity of the Hamiltonians $H_0$, $H_1$, and $H_2$.
If one introduces the new operators
\begin{equation}\label{redef}
 a_1=A_1, \hskip 3cm a_2=A_2^\dag
\end{equation}
the spectra of the Hamiltonians $H_0$, $H_1$ and $H_2$ are
related as follows
\begin{align*}
 \sigma(H_1)&\stackrel{a_1^\dag}\longrightarrow
  \sigma(H_0)\stackrel{a_2^\dag}\longrightarrow\sigma(H_2),&
 \sigma(H_1)&\stackrel{\,a_1}\longleftarrow
  \sigma(H_0)\stackrel{\,a_2}\longleftarrow\sigma(H_2).
\end{align*}
Let us introduce the following matrix operators
\begin{align}
 \hat H&=\begin{pmatrix}
       H_2&0&0\\0&H_0&0\\0&0&H_1
      \end{pmatrix},&
 Q&=\begin{pmatrix}
       0&0&0\\a_2&0&0\\0&a_1&0
      \end{pmatrix}.
\end{align}
These operators form the nonlinear algebra given by
\begin{align}\label{p2susy}
 [\hat H,\,Q]&=0, &
 Q^2Q^\dag + QQ^\dag Q + Q^\dag Q^2&=
 Q\bigl(P_1(\hat H) + P_2(\hat H)\bigr), &
 Q^3&=0,
\end{align}
and the hermitian conjugated relations. It is easy to see
that this algebra is a nonlinear version of the linear
parasupersymmetry of order $p=2$ \cite{rubakov}. This allows
to conclude that the one-dimensional quasi-exactly solvable
systems with two invariant subspaces are naturally related
not only to the nonlinear supersymmetry but to the nonlinear
parasupersymmetry as well.

\section{Supersymmetric quantum mechanics from viewpoint of
quasi-exact solvability}
\label{sqm}

As it was discussed in Section \ref{ir&qes} the intertwining
relations \eqref{Dt} underline the general construction of
one-dimensional supersymmetric quantum mechanics. If the
operator $A$ is a linear operator one arrives to the usual
supersymmetric quantum mechanics \cite{nicolai, witten}
while the construction with $A$ to be a higher order
differential operator corresponds to a nonlinear
generalization of the supersymmetry with a polynomial
superalgebra \cite{nsusy0, nsusy1, nsusy2, nsusy3, nsusy4}.
Owning to the discussed above correspondence between the
intertwining relations \eqref{Dt} and quasi-exact
solvability of one-dimensional systems we can consider the
supersymmetric quantum mechanics in this context.

In a general quantum mechanical system with a linear
supersymmetry the "bosonic" Hamiltonian can be treated as a
quasi-exactly solvable operator with a one-dimensional
invariant subspace. Since the sequence of eigenfunctions of
the Hamiltonian can always be reduced to the form
\eqref{space1} without loss of generality we can take the
invariant subspace in the form $\cF(H)=\{1\}$. It is easy to
see that if the second order differential operator is
invariant on this space then it has the following form:
\begin{equation}\label{H1}
 H={}-P(z)\pd_z^2 + Q(z)\pd_z + const.
\end{equation}
After changing the variable
$$
 x=\int\frac{dz}{\sqrt{P(z)}}
$$
and introducing the notation
$$
 W(x)=\left.\frac{P'(z)+2Q(z)}{4\sqrt{P(z)}}\right|_{z=z(x)}
$$
the Hamiltonian \eqref{H1} acquires the standard form
$$
 H={}-\pd_x^2+W(x)^2-W'(x) + const.
$$
One can note that while the operator \eqref{H1} depends on
two arbitrary functions the resulting operator of the
Schr\"odinger form counts on just one function, the
superpotential.

It is interesting to note that in this case the set of
annihilating operators is infinite-dimensional. Its basis
can be chosen in the form $L_n=z^{n+1}\pd_z$, $n\in\Z$,
hence, these operators form the Virasoro algebra without
central extension. The operator \eqref{H1} can be treated as
a combination of quadratic and linear terms in the
generators of the Virasoro algebra. One can conclude that
this fact is responsible for the presence of the functional
arbitrariness in the supersymmetric Hamiltonian.

Now consider the general case of a second order
quasi-exactly solvable operator with two-dimensional
invariant subspace. Without lost of generality one can deal
with the following subspace
\begin{align}\label{isp2}
 \cF(H)&=\mop{span}\left\{1,\,z\right\}.
\end{align}
The corresponding quasi-exactly solvable Hamiltonian can be
written as
\begin{align}\label{H2}
 H&={}-P(z)\pd_z^2 + \sum_{a=1}^3C_aJ^a_{1/2} + const,
\end{align}
where $C_a$ are constant and $J^a_{1/2}$ are linear
differential operators \eqref{sl2} forming a
representation of the algebra $\asl(2,\,\R)$ on the
two-dimensional space \eqref{isp2}.

The structure of the operator \eqref{H2} with the
two-dimensional invariant subspace is different as against
the previous case. The second term in \eqref{H2} is a linear
combination of affine annihilators of the invariant subspace
while the first one can be treated as a linear combination
of annihilators for that space.

The annihilators of the space \eqref{isp2} are any
differential operators of order 2 and higher without
linear in derivative and multiplicative terms.
Obviously, they form an infinite-dimensional set. The basis
of this set can be taken in the form
\begin{equation}\label{NVir}
 L_n=z^{n+2}\pd_z^2,\hskip 2cm n\in\Z.
\end{equation}
These operators together with the affine annihilators
$J^a_{1/2}$ generate the following nonlinear
infinite-dimensional algebra:
\begin{align}
 \bigl[L_m,L_n\bigr]&=(m-n)L_{m+n}\circ J^0_{1/2}, &
 \bigl[J^0_{1/2},L_m\bigr]&=m L_m,&
 \bigl[J^\mp_{1/2},L_m\bigr]&=(m\mp 2)L_{m\pm 1},&
\end{align}
where $\circ$ stands for anticommutator. Therefore, one can
conclude that the annihilators \eqref{NVir} together with
the affine annihilators form the nonlinear (quadratic)
generalization of the Virasoro algebra.

Using the usual intertwining relations \eqref{Dt} with the
annihilator $A=\pd_z^2$ one can reproduce the form of the
Hamiltonian \eqref{H2} and construct its superpartner. That
would complete the construction of the supersymmetric system
with a quadratic superalgebra \cite{andrian}.

Now let us investigate the case of a three-dimensional
invariant subspace, which can always be reduced to the
following form:
\begin{equation}\label{isp3}
 \cF(H)=\mop{span}\left\{1,\,z,\,f(z)\right\}.
\end{equation}
Because of the presence of the arbitrary real function
$f(z)$ this case looks different in comparison with the two
previous cases. As we shall see it is also different from
the viewpoint of quasi-exact solvability of second order
operators.

Using the usual intertwining relations \eqref{Dt} with the
annihilator of the subspace \eqref{isp3}
\begin{equation}\label{A3}
 A=\pd_z^3-\bigl(\log{f''(x)}\bigr)'\pd_z^2
\end{equation}
one can derive the form of corresponding Hamiltonian:
\begin{equation}\label{H3}
 H=\sum_{i,j=1}^3f_i(z)\,a_{ij}\,K_j,
\end{equation}
where the functions
\begin{align}\label{basis3}
 f_1(z)&=1,& f_2(z)&=z,& f_3(z)&=f(z)
\end{align}
form the basis of the invariant subspace \eqref{isp3} and
$a_{ij}$ are real numbers. The operators $K_i$ form a basis
of affine annihilators of the subspace \eqref{isp3}:
\begin{align*}
 K_1&=\I-zK_2-f(z)K_3,&
 K_2&=\pd_z-\frac{f'(z)}{f''(z)}\pd_z^2, &
 K_3&=f''(z)^{-1}\pd_z^2.
\end{align*}
They act on the basis \eqref{basis3} as follows:
$$
 K_i[f_j]=\delta_{ij}.
$$
From the viewpoint of the general structure \eqref{gen_qes}
of quasi-exactly solvable operators the Hamiltonian
\eqref{H3} is just a linear combination of affine
annihilators. Any annihilator of the invariant subspace
\eqref{isp3} is a differential operator of order not less
then 3. Therefore, the Hamiltonian cannot contain any of
them since we are restricted by the class of second order
differential operators.

The form of the supersymmetric partner of the Hamiltonian
\eqref{H3} can be derived from the corresponding
intertwining relations. We do not need to know it to obtain
the form of the superalgebra they generate together with the
supercharges (see \eqref{susy}). The conservation of
supercharges follows from the intertwining relations. The
anticommutator of supercharges is proportional to a cubic
polynomial in the super-Hamiltonian:
$$
 \left\{Q,\,Q^\dag\right\}=
 \mop{Det}\left(E\I-A\right)\bigr|_{E\to\hat H}
  =\hat H^3 +\ldots,
$$
where elements of the matrix $A=\|a_{ij}\|$ are defined
in \eqref{H3}. This follows form the fact that the
finite-dimensional invariant subspace \eqref{isp3} of the
``bosonic'' Hamiltonian \eqref{H3} corresponds to the
zero-mode space of the supercharge $Q$.

Construction of the cubic supersymmetry with Hamiltonians in
Schr\"odinger form was presented in Ref. \cite{sqm3}. Other
approaches to general quasi-exactly solvable operators with
three states can be found in Ref. \cite{3states}.

Now we would like to discuss the general case of
one-dimensional quasi-exactly solvable differential
operators of second order. If the operator \eqref{H} has a
$n$-dimensional invariant subspace \eqref{space0} its
coefficient functions and the basis functions $f_i(z)$ obey
the system of differential equations
\begin{equation}\label{gen}
 {}-P(z)f_i''(z) + Q(z)f_i'(z) + R(z)f_i(z)=
  \sum_{j=1}^nC_{ij}f_j(z),
\end{equation}
$i=1,\,\ldots,n$,
for some real constants $C_{ij}$. If $n\leq 3$ the equations
\eqref{gen} can be treated as a system of algebraic
equations for the variables $P(z)$, $Q(z)$, and $R(z)$. In
this manner the Hamiltonians \eqref{H1}, \eqref{H2}, and
\eqref{H3} can be reproduced. Beginning with $n=4$ the
situation changes drastically. Indeed, in the case $n=4$ the
system allows not only to fix the coefficient functions of
the Hamiltonian but it also leads to a nonlinear
differential equation for the functions $f_i(z)$.\footnote{
Or for the functions $f_3(z)$ and $f_4(z)$ in the case of
the reduced invariant subspace \eqref{space1}.}

Besides, from the system \eqref{gen} one can easily infer
that in the cases $n=1,\,2$ the functional arbitrariness of
the Hamiltonians is related to the degeneracy of the
corresponding algebraic system. The system \eqref{gen} also
allows to conclude that the nature of the functional
arbitrariness in the Hamiltonian \eqref{H3} is different
as against the cases \eqref{H1} and \eqref{H2}. For $n=1$
and $n=2$ the functional arbitrariness is related to the
degeneracy of the system of algebraic equations on the
coefficient functions while for $n=3$ it is related to the
presence of the arbitrary function in the basis
\eqref{space1}.

The Hamiltonian restricted to the invariant subspace is
represented by the matrix $C_{ij}$. Although it is real
valued but nonsymmetric matrix, hence, its eigenvalues can
be complex numbers. Therefore, if $C_{ij}$ is nonsymmetric
matrix one has to assume that the functions forming the
basic of the subspace do not have finite norm with respect
to the inner product \eqref{sp} since the Hamiltonian has to
be a hermitian operator. General structure of the matrix
$C_{ij}$ was studied in Ref. \cite{nsusy4} in the context of
polynomial superalgebras.

In the general context the most nontrivial problem is to
find and classify all functional finite-dimensional spaces
which can serve as invariant subspaces of differential
operators of the general form \eqref{H} and cannot be
reduced to the monomial form using changes of the variable
and gauge transformations. Such spaces do exist. For
example, consider the following one-parametric deformation
of the series \eqref{Fgap}:
\begin{align}\label{Fnew}
 \cF=\mop{span}\bigl\{1,\,z,\,z^2,\,\ldots,\,z^{n-3},
  \,z^{n-2}+\frac\alpha{n-1}z^{n-1},\,z^n\bigr\}
\end{align}
with $n\in\N$ and $\alpha\in\R$, which cannot be reduced to
the monomial form for $\alpha\ne 0$. The corresponding
annihilator of the subspace can be represented in the form
$$
 A_n=\bigl((\alpha z + 2)\pd_z-\alpha\bigr)
     (z\pd_z-2)\pd_z^{n-2}.
$$
Application of the method of intertwining relations leads to
a 6-parametric quasi-exactly solvable Hamiltonian which is
particular case of the $\asl(2,\R)$-based Hamiltonian with
$n\to n-2$ and the following constraints on the parameters:
\begin{align*}
 p_4&=0,&
 q_0&=\frac{n+2}2p_1 - \alpha p_0,&
 q_1&={}-\frac 2\alpha\left(n-1\right)p_3 + np_2
      + \alpha p_1 - \alpha^2p_0, &
 q_2&=\frac{n+4}2p_3.
\end{align*}

As another example of invariant subspaces of monomials
consider the following space:
\begin{align*}
 \cF=\mop{span}\bigl\{1,\,z,\,z^3,\,z^4+6\alpha z^2\bigr\}.
\end{align*}
Using the method of intertwining relations one calculates
the 7-parametric quasi-exactly solvable Hamiltonian given by
the coefficient functions
\begin{align*}
 P(z)&=p_5z^5 + p_3z^3 + p_2z^2+p_1z+p_0,\\
 Q(z)&=6p_5z^4 + 4\left(p_3+4\alpha p_5\right)z^2
     + \left(5p_2-\alpha^{-1}p_0\right)z
     - 8\alpha^2p_5 + 4\alpha p_3 + 2p_1,\\
 R(z)&={}-12p_5z^3 - 4\left(13\alpha p_5+p_3\right)z + r_0.
\end{align*}

More detailed discussion of quasi-exactly solvable operators
with non-monomial invariant subspaces will be presented
elsewhere.

\section{Conclusion}

In this paper we emphasized the method based on the
intertwining relations as a general approach to constructing
one-dimensional quasi-exactly solvable operators. We
demonstrated the universality of this method, which is based
on the algebraic nature of the intertwining relations. The
discussed examples cover all quasi-exactly solvable
operators with invariant monomial subspaces. It can be
equally applied to quasi-exactly solvable operators with
non-monomial invariant subspaces. Using the method we
constructed two examples of such operators with invariant
subspaces which cannot be reduced to a monomial space.

The intertwining relations themselves can be applied to the
one-dimensional case only. We offered natural algebraic
generalization that can be applied to the multidimensional
case as well. We showed that the simplest form of the
generalization allows to relate the quasi-exactly solvable
systems with two invariant monomial subspaces to a nonlinear
parasupersymmetry of second order. The application of the
generalized intertwining relations to constructing
quasi-exactly solvable operators in the multidimensional
case will be presented in future publications.

Besides we investigated quantum-mechanical systems both with
linear and nonlinear supersymmetry from the standpoint of
quasi-exact solvability. We demonstrated that in the cases
of linear and quadratic supersymmetry the functional
arbitrariness of the Hamiltonian is related to an
infinite-dimensional algebra of annihilators of
corresponding invariant subspaces. It is the Virasoro
algebra without central extension in the first case and a
quadratic deformation of the Virasoro algebra in the second.
Also we constructed a general quantum-mechanical system
associated with a cubic supersymmetry and argued that
quantum-mechanical systems with nonlinear supersymmetry of
order more then three are not quasi-exactly solvable in
general.


\end{document}